\documentclass[12pt]{article}
\usepackage{amsmath,amssymb}
\usepackage{graphicx}
\usepackage{amsmath}
\usepackage{hyperref}
\usepackage{float}
\usepackage{color, soul}
\sethlcolor{yellow}
\newcommand {\be}{\begin{equation}}
 \newcommand {\ee}{\end{equation}}
 \newcommand {\bea}{\begin{array}}
 
 \newcommand {\eea}{\end{array}}

\textwidth=6.5in \hoffset=-.75in \voffset=-1in \numberwithin{equation}{section}
\numberwithin{figure}{section}

\def\0{{(0)}}
\def\1{{(1)}}
\def\2{{(2)}}

\def\<{\langle }
\def\>{\rangle }
\def\[{\left[}
\def\]{\right]}

\begin{document}
\begin{titlepage}

\vskip1cm
\begin{center}
{~\\[140pt]{ \LARGE {\textsc{Holographic Entanglement negativity in flat space generalized minimal massive gravity    }}}\\[-20pt]}
\vskip2cm

\end{center}
\begin{center}
{Mohammad Reza Setare \footnote{E-mail: rezakord@ipm.ir }\hspace{1mm} ,
 Meisam Koohgard \footnote{E-mail: m.koohgard@modares.ac.ir}\hspace{1.5mm} \\
{\small {\em  {Department of Science,\\
 Campus of Bijar, University of Kurdistan, Bijar, Iran }}}}\\
\end{center}
\begin{abstract}
 In this paper we study the application of holographic entanglement negativity proposal for bipartite states in the 2d Galilean conformal field theory ($GCFT_2$) dual to bulk asymptotically flat spacetimes in the context of generalized minimal massive gravity (GMMG) model. $GCFT_2$ is considered on the boundary side of the duality and the bulk gravity is described by GMMG that is asymptotically symmetric under the Galilean conformal transformations. In this paper, the replica technique, based on the two-point and the four-point twist correlators, is utilized and the entanglement entropy and the entanglement negativity are obtained in the bipartite configurations of the system in the boundary.         \end{abstract}
\vspace{1cm}


\end{titlepage}

\section{Introduction}	
Recently, the non-relativistic version of the duality between gravity
theory and boundary quantum field theory has attracted a lot of attentions. One of the areas that this type of gauge/gravity duality can be interesting is the real-life systems in condensed
matter physics. The Schr\"{o}dinger group as a non-relativistic conformal symmetry group has been used in studying the cold atoms \cite{CM1,CM2,CM3}. This symmetry can be realized by taking the non-relativistic limit of the Klein-Gordon equation in the massive case. The Galilean conformal group (GC), which arises as parametric contraction of conformal group, is different with Schr\"{o}dinger group and has been used to study the quantum critical behaviors in a family of universality classes \cite{gp1}. Especially this symmetry governs the behavior of the ultra-cold
atoms at nucleon scattering in some channels \cite{CM1,gp3}.

 In this paper, we consider the Galilean conformal group. The Galilean conformal algebra (GCA) was explored in two dimensions by
two copies of Virasoro algebra in non-relativistic limit \cite{gal3}. An interesting feature of GCA is related to its extension to an infinite dimensional symmetry algebra \cite{gal3}.
One option
for bulk description of 2d GCA, with asymmetric central charges would be
generalized minimal massive gravity (GMMG) in three dimensions \cite{Setare1}. GMMG model providing a new example of a
theory that avoids the bulk-boundary clash and therefore, as Minimal Massive Gravity (MMG) \cite{mmg}, the theory possesses both, positive energy excitations around the maximally $AdS_3$ vacuum
as well as a positive central charge in the dual conformal field theory (CFT). Such clash is present in the previously constructed
gravity theories with local degrees of freedom in 2+1-dimensions, namely Topologically
Massive Gravity (TMG) \cite{Des01,Des02} and the cosmological extension of New Massive Gravity (NMG) \cite{nmg}. MMG supplements the field equations
of Topologically Massive Gravity with a symmetric, ranked two-tensor containing up to second
derivatives of the metric. GMMG model is a general version of MMG in which by adding the CS deformation term (stands for Chern-Simons term
for the one-form Levi-Civita affine connection which produces a propagating massive graviton),
and an extra term to pure Einstein gravity with a negative cosmological constant. Since these theories avoid the bulk-boundary clash, they define excellent arenas to explore
the structure of asymptotically AdS solutions, asymptotic symmetries, their algebra and other
holographically inspired questions. We study some aspects of a class of non-AdS holography where the $3d$ bulk gravity is given by GMMG. We investigate a derivation of the holographic entanglement negativity using $flat_3/GCFT_2$ in the context of GMMG.\\

The study of the quantum entanglement of many-body systems has increased the knowledge of these systems (see for a review \cite{qE01}). The quantum entanglement has an important role to study of a wide variety of the fields in physics from the condensed matter systems to quantum gravity. One of the most notable measures of quantum entanglement is the entanglement entropy \cite{qE02}. The entanglement entropy can be defined by the von Neumann
entropy of the corresponding reduced density matrix of a subsystem of a bipartite quantum system. When the whole system is in a state $|\Psi\rangle$ with the density matrix $\rho=|\Psi\rangle\langle\Psi|$, the reduced density matrix of A is $\rho_A=\mathrm{Tr}_B\rho$. The observer A can observe only the subset of a complete set of commuting observables. The other observer B can observe the remainder of the system. Using the von Neumann entropy, the entanglement entropy is $S_A=-\mathrm{Tr}_A\rho_A\log\rho_A$. This approach can not give a proper measure about the entanglement between two non-complementary subsections in a mixed state \cite{repl1}.

Some logarithmic measures of the entanglement in the bipartite states has been introduced in \cite{plen1,plen2,mes1,mes2}, but these measures have some problems to be evaluated analytically. Another measure of the entanglement has been introduced in \cite{Vida} (see also \cite{Vida2}) that is a computable one. This measure is the \emph{entanglement negativity} that can be obtained from any mixed state. A classical approach to compute the negativity in quantum field theory (QFT) and many-body systems has been introduced in \cite{repl1,repl2} . This method provides a good measure for entanglement between two parts of a larger system in a pure state and has been used a lot to study the ground-states of CFTs \cite{cft1,cft2,cft3,cft4}. The authors in \cite{Malv} extended the works of \cite{repl3,repl4} on the entanglement entropy of GCFTs to compute the entanglement negativity of the $GCFT_2$. The replica technique has been used to compute the entanglement entropy of the bipartite configurations of $GCFT_2$ in \cite{Malv}. It can be seen that the universal features of the entanglement negativity in $CFT_2$ can be reproduced in $GCFT_2$ in \cite{Malv}. In \cite{Basu}, a covariant construction in calculation of the entanglement negativity has been introduced in the context of bipartite states in the duality $GCFT_2$/TMG where this is in a class of flat-holography. It can be seen that the universal part of the entanglement negativity is same as the universal part of the negativity calculated in \cite{Malv}.

In this paper, we extend the measurement of the quantum entanglement to another duality in the flat holography that the bipartite system can be considered in the boundary side of the duality and the GMMG is the gravity theory in the bulk. To compute the entanglement entropy and the entanglement negativity utilizing the replica technique, it is crucial to find out the two-point and four-point twist correlators for the bipartite configuration of the subsystem. We find a viable form of the correlators in the $GCFT_2$/GMMG based on the construction of the entanglement entropy in the bulk geometry with a conical singularity that the result can be reproduced by comparing with the result in our previous work \cite{Setare3}. The equality of the universal features of the entanglement negativity in this work matches with the universal parts of the negativity in \cite{Malv}.

This paper is organized as follows. In section \ref{sec:2} the general features of the GCA is presented.
We give the essential tools that are needed to utilize the replica technique in this context. We present the argument to find out the two-point twist correlator in the GMMG by asymptotically flat geometry, in section \ref{sec:3}. The first case we consider is related to a single interval configurations. These configurations are considered in section \ref{sec:4} on the zero and finite temperature and on the finite-sized systems. The bipartite configurations with two adjacent intervals and with two disjoint intervals are considered in sections \ref{sec:5} and \ref{sec:6}, respectively. We conclude in section \ref{sec:7} with a summary of our results.

\section{Replica approach in $GCFT_2$}\label{sec:2}
The replica technique has been used to compute the entanglement entropy and the entanglement negativity in some special cases of the flat holography. Also this technique has been used to compute the entanglement negativity in a $CFT_2$ \cite{repl1,repl2}. Another application of the replica technique can be found in the computation of the entanglement entropy in Galilean conformal field theories ($GCFT_2$) in \cite{repl3,repl4}. The two dimensional Galilean conformal field theories \cite{gal1,gal2,gal3,gal4} are in a class of conformal field theories with Galilean invariance. The $GCFT_2$ involves the Galilean conformal algebra (GCA) that may be obtained via usual conformal algebra in two dimensions as follows
\begin{equation}\label{Gal.Tr}
  t\to t,~~~x\to \epsilon x ,
\end{equation}
with $\epsilon\to 0$. The generators of $GCA_2$ in the plane coordinates are given by the following relations
\begin{equation}\label{Gal.gen}
  L_n=t^{n+1}\partial_t+(n+1)t^nx\partial_x,~~~ M_n=t^{n+1}\partial_x,
\end{equation}
where these generators satisfy the following Lie algebra with a central extension
\begin{eqnarray}\label{Gal.alg}
  \[L_n,L_m\] &=& (m-n)L_{n+m}+\frac{c_L}{12}(n^3-n)\delta_{n+m},\nonumber \\
  \[L_n,M_m\] &=& (m-n)M_{n+m}+\frac{c_M}{12}(n^3-n)\delta_{n+m},\nonumber \\
  \[M_n,M_m\] &=& 0,
\end{eqnarray}
where $c_L$ and $c_M$ are the central charges of the algebra $GCA_2$. To compute the entanglement entropy, the replica manifold can be considered as an \emph{n}-sheeted surface. This surface involves \emph{n} copies of the $GCFT_2$ plane glued together.
The reduced density matrix $\rho_A^n$ corresponding to the interval $A$ can be considered as follows \cite{repl1}
\begin{equation}\label{red.M01}
  Tr \rho_A^n=\langle \Phi_n(\partial_1A)\Phi_{-n}(\partial_2A)\rangle,
\end{equation}
where the twist fields $\Phi_n$ are primary fields of the $GCFT_2$ with scaling dimensions as follows\footnote{The scaling dimensions $\Delta_n$ and $\chi_n$ are related to the weights $h_L$ and $h_M$ of the twist fields by the following relations that have been introduced in \cite{repl3,repl4}

\begin{equation*}
  \Delta_n=n h_L,~~~~\chi_n=n h_M.
\end{equation*}  }
\begin{equation}\label{DK01}
  \Delta_n=\frac{c_L}{24}(n-\frac{1}{n}),~~~\chi_n=\frac{c_M}{24}(n-\frac{1}{n}).
\end{equation}

Now, the entanglement entropy for the bipartite state corresponding to the interval $A$ can be computed by taking the replica limit as follows \cite{repl1}
\begin{equation}\label{EE01}
  S_A=-\lim_{n\to 1}\frac{\partial}{\partial n}\mathrm{Tr}\rho_A^n.
\end{equation}

To define the entanglement negativity as \cite{Vida}, a tripartite system is considered to consist of the subsystems $A_1$, $A_2$ and $B$ where $A=A_1\cup A_2$ and $B=A^c$ represent the rest of the system. The entanglement negativity for the bipartite mixed state $A=A_1\cup A_2$ can be defined as follows
\begin{equation}\label{EN01}
  \mathcal{E}=\log \mathrm{Tr}\| \rho_A^{T_2}\|,
\end{equation}
where $\rho_A^{T_2}$ is the partial transpose of the reduced density matrix for the system $A$ with respect to  the susbsystem $A_2$ that can be defined as follows
\begin{equation}\label{red.M02}
  \big\langle e_i^{(1)}e_j^{(2)}|\rho_A^{T_2}|e_k^{(1)}e_l^{(2)}\big\rangle=
  \big\langle e_i^{(1)}e_l^{(2)}|\rho_A|e_k^{(1)}e_j^{(2)}\big\rangle,
\end{equation}
where $|e_i^{(1)}\rangle$ and $|e_j^{(2)}\rangle$ are the bases for the Hilbert spaces $\mathcal{H}_1$ and $\mathcal{H}_2$. The replica technique involves an analytic continuation of the reduced density matrix through even or odd sequences of $n=n_e=2m$ or $n=n_o=2m+1$, respectively\cite{repl1}. Because of the non-trivial form of the quantity $\mathrm{Tr}( \rho_A^{T_2})$  through even sequences, this type of the analytic continuation is more desired.
The partial transpose of the reduced density matrix $\rho_A^n$ corresponding to the interval $A$ through the $n$-even sequences can be used  to compute the entanglement negativity $\mathcal{E}$ for the bipartite mixed state $A$ as follows \cite{Malv}

\begin{equation}\label{EN02}
  \mathcal{E}=\lim_{n\to 1}\log \mathrm{Tr}( \rho_A^{T_2})^{n},
\end{equation}
where we have used the replica limit $n\to 1$. The quantity $\mathrm{Tr}( \rho_A^{T_2})^{n}$ can be expressed as follows
\begin{equation}\label{Q01}
  \mathrm{Tr}( \rho_A^{T_2})^{n}=\langle\Phi_n(x_1,t_1)\Phi_{-n}(x_2,t_2)\Phi_{-n}(x_3,t_3)\Phi_n(x_4,t_4) \rangle.
\end{equation}

\section{Two-point twist correlator in GMMG }\label{sec:3}
The holographic entanglement entropy for a single interval in the vacuum state of a $GCFT_2$ can be done by the replica technique. To this end, the geometry of the bulk should be considered. So far, this technique has been utilized in the Einstein gravity and in the flat space topological massive gravity (TMG) \cite{Basu}. At first we review the results of \cite{Basu} briefly, then we apply the replica approach in the generalized minimal massive gravity (GMMG) \cite{Setare1} in this section. By comparing the result in the GMMG to the TMG result in the entanglement entropy, we find that the effect of GMMG on the use of replica technique only can be seen in changing the central charges\footnote{The $GCFT_2$'s which are dual with GMMG and TMG have a distinction on their central charges. This difference can be found in the following, but this is not the only distinction. The difference between the central charges changes the eigenvalues and the conformal weights of the twist fields in GMMG and TMG duals. Another difference we find between $GCFT$'s dual with GMMG and TMG is the change in the correlation functions, and these correlation functions play an important role in obtaining entanglement ingredients. Since the correlation functions are also obtained by the expansion of conformal blocks, another distinction between dual $GCFT$'s of TMG and GMMG is the distinction between conformal blocks. }, and this is a significant result. This becomes more interesting due to the existence of the higher derivative term in the GMMG Lagrangian.

 In this context, the flat space geometry in Eddington-Finklestein coordinates is given by the following metric
\begin{equation}\label{glob.met}
  ds^2=dr^2-du^2+r^2d\phi^2.
\end{equation}

An interval $A=[(u_1,\phi_1),(u_2,\phi_2)]$ on the boundary plane is located at the infinity of the flat spacetime. The length of the bulk extremal curve joining the endpoints of the interval $A$ is given by \cite{Hij01}
\begin{equation}\label{L01}
  L_{tot}^{extr}=\bigg |\frac{u_{12}}{\tan\frac{\phi_{12}}{2}} \bigg |,
\end{equation}
where
\begin{equation}\label{uP}
  u_{12}=u_1-u_2,~~~\phi_{12}=\phi_1-\phi_2.
\end{equation}

The curve consists of two null geodesics and another spacelike one. The null geodesics have been extended from the interval endpoints and the spacelike geodesic connects two null curves. Since the null geodesics have zero length, the extremal length (\ref{L01}) is obtained by extremizing the spacelike geodesic that connects them. For the Einstein gravity in the flat space we have the following central charges for dual $GCFT_2$
\begin{equation}\label{cLM01}
  c_L=0,~~~c_M=\frac{3}{G}.
\end{equation}

Substituting these central charges into (\ref{DK01}), the field scaling dimensions can be found as follows
\begin{equation}\label{DK02}
  \Delta_n=0,~~~\chi_n=\frac{1}{8G}(n-\frac{1}{n}).
\end{equation}

As described in \cite{Hij01}, the twist fields inserted at the endpoints of the interval can be considered as a particle with mass $m_n=\chi_n$ that propagates in the bulk. The two point correlator for the twisted fields can be considered as follows \cite{Hij01}
\begin{equation}\label{Q02}
  \langle\Phi_n(\partial_1A)\Phi_{-n}(\partial_2A)\rangle=\exp(-\chi_nS_{on-shell}),
\end{equation}
where
\begin{equation}\label{S01}
  S_{on-shell}=\sqrt{\eta_{\mu\nu}\dot{X}^{\mu}\dot{X}^{\nu}}=L_{tot}^{extr},
\end{equation}
where $S_{on-shell}$ is the on-shell action of the particle that propagates along an trajectory $X^{\mu}$
homologous to the interval. Substituting (\ref{S01}), (\ref{DK02}) and (\ref{L01}) into (\ref{Q02}), one can find the two-point correlator as follows
\begin{equation}\label{Q03}
  \langle\Phi_n(\partial_1A)\Phi_{-n}(\partial_2A)\rangle=\exp\bigg(-\frac{1}{8G}(n-\frac{1}{n})
  \bigg |\frac{u_{12}}{\tan\frac{\phi_{12}}{2}} \bigg |\bigg),
\end{equation}

Using the replica limit (\ref{EE01}), and Eqs.(\ref{red.M01}), (\ref{Q03}) the entanglement entropy can be found as follows
\begin{equation}\label{EE02}
  S_A=\frac{1}{4G}L_{tot}^{extr}=\frac{1}{4G}\bigg |\frac{u_{12}}{\tan\frac{\phi_{12}}{2}} \bigg |.
\end{equation}

By considering the TMG in flat space \cite{Des01,Des02}, a non-vanishing $\Delta$
introduces a spin for the massive particle. In this context, the bulk picture can be modified by  introducing a gravitational Chern-Simons (CS) term \cite{Basu,Hij01}. The action of the TMG has the following form
\begin{eqnarray}\label{TMG.S}
  \mathcal{S}_{TMG} &=& \mathcal{S}_{EH}+\frac{1}{\mu}\mathcal{S}_{CS} \nonumber \\
   &=& \frac{1}{16\pi G}\int d^3x \bigg (R+\frac{2}{l^2}+\frac{1}{2\mu}
   \epsilon^{\alpha\beta\gamma}\big(\Gamma^{\rho}_{\alpha\sigma}\partial_{\beta}\Gamma^{\sigma}_{\gamma\rho}
   +\frac{2}{3}\Gamma^{\rho}_{\alpha\sigma}\Gamma^{\sigma}_{\beta\eta}\Gamma^{\eta}_{\gamma\rho}\big)\bigg),
\end{eqnarray}
where the first term is the usual Einstein-Hilbert term, the second term is the cosmological constant term and the last one is a gravitational CS term. The algebra of the asymptotic Killing vectors of the TMG in $AdS_3$ is isomprphic to two copies of the Virasoro algebra with the following right and left cental charges \cite{Brown}:
\begin{equation}\label{Vir.cen}
  c_{TMG}^+=\frac{3l}{2G}(1+\frac{1}{\mu l}),~~~c_{TMG}^-=\frac{3l}{2G}(1-\frac{1}{\mu l}).
\end{equation}

By taking Inonue-Wigner contraction \cite{Ino,bagG}, the asymptotic symmetry group leads to the Galilean conformal algebra (GCA) with the following central charges:
\begin{eqnarray}\label{cLM-TMG}
  c_L &=& c_{TMG}^+-c_{TMG}^-=\frac{3}{\mu G}, \nonumber \\
  c_M &=& \frac{(c_{TMG}^++c_{TMG}^-)}{l}=\frac{3}{G}.
\end{eqnarray}

In \cite{Basu}, the authors computed the entanglement entropy of the flat-TMG. As easily can be seen from (\ref{TMG.S}), the CS-term is unaffected by the flat limit $l\to\infty$ and the propagating particle along the bulk trajectory is a massive spinning one with mass $m=\chi_n$ and spin $s=\Delta_n$. For such a particle propagating along the trajectory homologous to the interval, the two-point twist correlator has the following form:
\begin{equation}\label{Q04}
  \langle\Phi_n(\partial_1A)\Phi_{-n}(\partial_2A)\rangle=\exp(-\chi_nS_{on-shell}^{EH}-\Delta_nS_{on-shell}^{CS}),
\end{equation}
where $S_{on-shell}^{CS}$ has the following form in a pure Minkowski spacetime \cite{Hij01}
\begin{equation}\label{eta.Min}
  S_{on-shell}^{CS}\equiv \Delta \eta_{12}=2\log\big(\frac{2}{\epsilon}\sin\frac{\phi_{12}}{2}\big).
\end{equation}
where $\Delta \eta_{12}$ is the extremized boost that is needed to take the parallel transport the normal vectors on the interval endpoints along the aforementioned geodesics.

Using the replica limit (\ref{EE01}), (\ref{red.M01}),(\ref{DK01}),(\ref{Q04}), the entanglement entropy for a single interval in a pure Minkowski
spacetime can be obtained as follows
\begin{eqnarray}\label{EE-T01}
  S_A &=& -\lim_{n\to 1}\frac{\partial}{\partial_n}\langle\Phi_n(\partial_1A)\Phi_{-n}(\partial_2A)\rangle \nonumber\\
   &=& \lim_{n\to 1}\big[\partial_n\chi_nS_{on-shell}^{EH}+\partial_n\Delta_nS_{on-shell}^{CS} \big] \nonumber\\
   &=& \frac{c_M}{12}S_{on-shell}^{EH}+\frac{c_L}{12}S_{on-shell}^{CS}.
\end{eqnarray}
Using (\ref{S01}) and substituting (\ref{L01}), (\ref{eta.Min}) into (\ref{EE-T01}), the following result can be found
\begin{equation}\label{EE-T02}
  S_A=\frac{c_M}{12}\bigg |\frac{u_{12}}{\tan\frac{\phi_{12}}{2}} \bigg |+\frac{c_L}{6}\log\big(\frac{2}{\epsilon}\sin\frac{\phi_{12}}{2}\big).
\end{equation}

The approach presented above is based on the geometry and the physics in the TMG. We show that the two-point correlator (\ref{Q04}) can also be used for the GMMG model, only the appropriate twist fields scaling dimensions and the right central charges must be placed. Then we find that the entanglement entropy that we get with this assumption is completely consistent with the result in \cite{Setare3}, and we construct this to use two-point correlator in the negativity computations as well. This result can be significant, given the differences between the two models TMG and GMMG. In this step, more details of the GMMG are provided. The Lagrangian of the GMMG model has the following form \cite{Setare1}
\begin{equation}\label{GM01}
  L_{GMMG}=L_{TMG}-\frac{1}{m^2}\big(f.R+\frac{1}{2}e.f\times f\big)+\frac{\alpha}{2}e.h\times h
\end{equation}
 where $L_{TMG}$ is the Lagrangian of the TMG and $R$ is the Ricci scalar. $f$ is an auxiliary one-form field and $m$ is a mass parameter of NMG term \cite{nmg}. $e$  is a dreibein and $h$ is the auxiliary field. The CS-term in the $L_{TMG}$ has the following form in the first order formalism
 \begin{equation}\label{CS01}
   CS-term: \frac{1}{2\mu}(\omega.d\omega+\frac{1}{3}\omega.\omega\times\omega)
 \end{equation}
where $\omega$ is dualised spin-connection. The algebra of the asymptotic conserved charges of asymptotically $AdS_3$ spacetimes in the context of GMMG is isomorphic to two copies of the Virasoro algebra with the following right and left central charges \cite{Setare1}
\begin{equation}\label{Vir.cen2}
  c_{GMMG}^+=\frac{3l}{2G}(-\sigma-\frac{\alpha H}{\mu}-\frac{F}{m^2}+\frac{1}{\mu l}),~~~c_{GMMG}^-=\frac{3l}{2G}(-\sigma-\frac{\alpha H}{\mu}-\frac{F}{m^2}-\frac{1}{\mu l}),
\end{equation}
where $H$ and $F$ are two constants. By taking Inonue-Wigner contraction, we can find the asymptotic symmetry group as a
 GCA with the following central charges
 \begin{eqnarray}\label{cLM-GMMG}
  c_L &=& c_{GMMG}^+-c_{GMMG}^-=\frac{3}{\mu G}, \nonumber \\
  c_M &=& \frac{(c_{GMMG}^++c_{GMMG}^-)}{l}=\frac{3}{G}(-\sigma-\frac{\alpha H}{\mu}-\frac{F}{m^2}).
\end{eqnarray}

\subsection*{The two point correlation function in the GMMG case}
The GMMG model is one of the models with gravitational anomalies in which the Ryu-Takayanagi (RT) successful conjecture needed some considerations \cite{RT01,RT02,RT03}. In the RT prescription to find out the holographic entanglement entropy, the action of the bulk gravity can be considered by the path traced out by a massive particle, but this picture should be modified by the spinning feature of the particle.

To find the holographic entanglement entropy using the replica technique, it is needed to define the gravitational partition functions that include the 3d bulk manifold. The authors in \cite{Mald,Cas14} consider the bulk action with a conical singularity of $2\pi\epsilon$ along a worldline $C$ in the bulk where $\epsilon$ is equal $n-1$. $n$ defines the number of the copies of the plane that are sewn together to form the boundary interval in the holography.
The entanglement entropy can be found by the conical action using the following relation  \cite{Cas14}
\begin{equation}\label{SC01}
  S_A=-\partial_{\epsilon}\big(S_{cone}    \big)|_{\epsilon=0}.
\end{equation}

In the Einstein-Hilbert gravity, the curve $C$ that is connected to the two ends of the entanglement interval $A$ at the boundary, should be a geodesic and the entanglement entropy result is consistent with the Ryu-Takayanagi (RT) conjecture as follows \cite{Cas14,RT01,RT02,RT03}
\begin{equation}\label{SEH}
  S_A^{(EH)}=\frac{A_{min}}{4G},
\end{equation}
where $A_{min}$ is the area of the bulk entangling surface that is connected to the two ends of the entanglement interval $A$.

Now, the cone metric can be considered as follows \cite{Cas14}
\begin{eqnarray}\label{mCo}
  ds^2 &=& e^{\epsilon\phi(\gamma)}\delta_{ab}d\gamma^a d\gamma^b+(g_{zz}+K_a\gamma^a+...)dz^2 \nonumber\\
  && + e^{\epsilon\phi(\gamma)}U_a(\gamma,z)d\gamma^a dz,
\end{eqnarray}
where $z$ is a spacelike direction along the cone worldline and $\gamma^a$ are two perpendicular flat coordinates. $K_a$ as extrinsic curvatures are the expansion coefficients of $g_{zz}$ along $\gamma^a$ directions. $U_a$ are some arbitrary functions and $\phi(\gamma)$ is a function that is needed to regularize the cone. By the technique extended in \cite{Mald,Cas14,Dong,Camp}, the regularized cone action in the bulk that is decribed by the TMG can be found as follows
\begin{eqnarray}\label{SCT01}
  S_{cone}|_{TMG} &=& -\frac{\epsilon}{4G} \int_{C}dz\sqrt{g_{zz}}\nonumber\\
  && -\frac{i\epsilon}{16\mu G}\int_{C}dz\epsilon^{ab}\partial_a U_b+\mathrm{O}(\epsilon^2)
\end{eqnarray}
that by considering the GMMG modifications in the TMG action, we find the regularized cone action in the GMMG as follows
\begin{eqnarray}\label{SCO01}
  S_{cone}|_{GMMG} &=& -\frac{\epsilon}{4G} (-\sigma-\frac{\alpha H}{\mu}-\frac{F}{m^2}) \int_{C}dz\sqrt{g_{zz}}\nonumber\\
  && -\frac{i\epsilon}{16\mu G}\int_{C}dz\epsilon^{ab}\partial_a U_b+\mathrm{O}(\epsilon^2)
\end{eqnarray}
where the action have been computed to the first order in $\epsilon$ in the expansion. The second terms in (\ref{SCT01}) and (\ref{SCO01}) are related to the CS-terms in the TMG and the GMMG Lagrangian, respectively. These terms have the same forms in the conical actions of these models.

The first term in (\ref{SCO01}) can be written in a covariant manner as follows
\begin{eqnarray}\label{SCO02}
  S_{cone}|_{GMMG} &=& -\frac{\epsilon}{4G} (-\sigma-\frac{\alpha H}{\mu}-\frac{F}{m^2}) \int_{C}d\tau \sqrt{g_{\mu\nu}\big(X(\tau)\big)\dot{X}^{\mu}\dot{X}^{\nu}   }\nonumber\\
  && -\frac{i\epsilon}{16\mu G}\int_{C}dz\epsilon^{ab}\partial_a U_b+\mathrm{O}(\epsilon^2)
\end{eqnarray}
where $X^{\mu}(\tau)$ parameterizes the cone tip and $\dot{X}^{\mu}\equiv \frac{\partial X^{\mu}}{\partial\tau}$.

Defining the following normal vectors
\begin{eqnarray}\label{nvs}
  n_{1} &\equiv& \frac{\partial}{\partial\gamma^{1}},  \nonumber\\
  n_{2} &\equiv & \frac{\partial}{\partial\gamma^{2}},
\end{eqnarray}
the $CS$-term in the $S_{cone}$ can be modified and we have the following relation
\begin{eqnarray}\label{SCO03}
  S_{cone}|_{GMMG} &=& -\frac{\epsilon}{4G} (-\sigma-\frac{\alpha H}{\mu}-\frac{F}{m^2}) \int_{C}d\tau \sqrt{g_{\mu\nu}\big(X(\tau) \big)\dot{X}^{\mu}\dot{X}^{\nu}  }\nonumber\\
  && -\frac{i\epsilon}{4\mu G}\int_{C}d\tau n_{2}.\nabla n_{1}+\mathrm{O}(\epsilon^2),
\end{eqnarray}
where the "$\nabla$" symbol indicates the covariant derivative along the worldline.

Utilizing (\ref{SC01}) to find the entanglement entropy, we find the full relation of the entanglement entropy as follows
\begin{eqnarray}\label{SA01}
  S_{A} &=& \frac{1}{4G} (-\sigma-\frac{\alpha H}{\mu}-\frac{F}{m^2}) \int_{C}d\tau \sqrt{g_{\mu\nu}\big(X(\tau)\big)\dot{X}^{\mu}\dot{X}^{\nu}   }\nonumber\\
  && +\frac{1}{4\mu G}\int_{C}d\tau \tilde{n}_{2}.\nabla \tilde{n}_{1}+\mathrm{O}(\epsilon^2),
\end{eqnarray}
where we have changed the Euclidean signature to the Lorentzian one, as follows
\begin{eqnarray}\label{EuLo}
  \tilde{n}_{1} &\equiv& i n_{1}=\partial_t, \\
  \tilde{n}_{2} &\equiv& n_{2}.
\end{eqnarray}

Using the GMMG central charges (\ref{cLM-GMMG}) in (\ref{SA01}), we find the following relation
\begin{equation}\label{SA02}
  S_{A} = \frac{c_M}{12}\int_{C}d\tau \sqrt{g_{\mu\nu}\big(X(\tau)\big)\dot{X}^{\mu}\dot{X}^{\nu}   }
  +\frac{c_L}{12}\int_{C}d\tau \tilde{n}_{2}.\nabla \tilde{n}_{1}+\mathrm{O}(\epsilon^2).
\end{equation}

Substituting the entanglement entropy (\ref{SA02}) into (\ref{EE01}), the two point correlation function (\ref{Q04}) can be reproduced in the GMMG with appropriate scaling dimensions.

We find the only change in the two-point twist correlator (\ref{Q04}) compared to the TMG is a change in the scaling dimension of the fields.  The effects of the GMMG Lagrangian on the two-point twist correlator (\ref{Q04}) can be addressed into the scaling dimensions $\chi_n$ and $\Delta_n$ compared to the flat-TMG. To provide another evidence for this result, we consider the two point correlation function of the twist field at the interval endpoints as follows
 \begin{eqnarray}\label{QG04}
  \langle\Phi_n(\partial_1A)\Phi_{-n}(\partial_2A)\rangle &=& \exp\bigg(- \frac{1}{8G} (-\sigma-\frac{\alpha H}{\mu}-\frac{F}{m^2})(n-\frac{1}{n})  S_{on-shell}^{EH}\nonumber\\
  &&-\frac{1}{8\mu G}(n-\frac{1}{n}) S_{on-shell}^{CS}\bigg).
\end{eqnarray}

 By using (\ref{S01}) and (\ref{eta.Min}) into (\ref{QG04}) and applying the replica technique (\ref{EE-T01}),
 we find the entanglement entropy for the single interval $A$ in the pure Minkowski spacetime as follows
 \begin{equation}\label{EE-G01}
  S_A=\frac{1}{4G}(-\sigma-\frac{\alpha H}{\mu}-\frac{F}{m^2})
  \bigg |\frac{u_{12}}{\tan\frac{\phi_{12}}{2}} \bigg |+\frac{1}{2\mu G}\log\big(\frac{2}{\epsilon}\sin\frac{\phi_{12}}{2}\big).
\end{equation}

This result is equal to the entanglement entropy for the GMMG that we have calculated in the flat holography in \cite{Setare3} by the Rindler transformation approach. This could be a sign that the replica technique, as we have used it, could be a true generalization to the GMMG one. In other words, obtaining the two-point correlation functions that are important in the calculations of the negativity can be the same as we have done here.

It should also be added that we have considered the Bondi-Metzner-Sachs ($BMS_3$) symmetry group \cite{Bondi1,Sa1,Sa2}  as an asymptotically symmetry group of GMMG in flat space in \cite{Setare3}. Since the $BMS_3$ field theories and the $GCFT_2$ theories have the same symmetry algebra, we expect the entanglement entropy to be the same in both cases. We arrive at this conclusion, and the entropy (\ref{EE-G01}) obtained from (\ref{Q04}) with the help of the replica approach is the same as the universal part of the result in \cite{Setare3}. The proof presented in this section shows us that we can use (\ref{Q04}) with (\ref{cLM-GMMG}) in the all correlators that we need in all related computations to the entanglement entropy and the entanglement negativity.

\section{Entanglement negativity and entanglement entropy for a single interval  }\label{sec:4}
In this section, we compute the entanglement negativity for various state configurations in a $GCFT_2$. These configurations consist of a single interval $A=[(x_1,t_1),(x_2,t_2)]$. To derive the entanglement negativity we need the four-point twist correlator (\ref{Q01}) that is considered for a tripartite system. In the bipartite limit $B\to\emptyset$, the four-point correlator (\ref{Q01}) can be turned into a two-point correlator as follows
\begin{equation}\label{EQ01}
  \mathrm{Tr}( \rho_A^{T_2})^{n}=\langle\Phi^2_n(x_1,t_1)\Phi^2_{-n}(x_2,t_2) \rangle.
\end{equation}

For even $n$, there is following relation between the $n$-sheeted surface correlator and the $n/2$-sheeted surface correlator \cite{repl2}
\begin{equation}\label{EQ02}
  \langle\Phi^2_n(x_1,t_1)\Phi^2_{-n}(x_2,t_2) \rangle=(\langle\Phi_{n/2}(x_1,t_1)\Phi_{-n/2}(x_2,t_2) \rangle)^2
\end{equation}

Substituting (\ref{Q04}) into the left-hand side of (\ref{EQ02}), we find the two-point correlator (\ref{EQ01}) as follows
\begin{equation}\label{EQ03}
  \langle\Phi^2_n(\partial_1A)\Phi^2_{-n}(\partial_2A)\rangle=\exp(-\chi_{n/2}
  S_{on-shell}^{EH}-\Delta_{n/2}\Delta\eta_{12}),
\end{equation}
where
\begin{eqnarray}\label{cLM-GMMG2}
  \chi_{n/2} &=& \frac{c_M}{24}\big(\frac{n}{2}-\frac{2}{n}\big)=
  \frac{1}{8G}(-\sigma-\frac{\alpha H}{\mu}-\frac{F}{m^2})\big(\frac{n}{2}-\frac{2}{n}\big) \nonumber \\
  \Delta_{n/2} &=& \frac{c_L}{24}\big(\frac{n}{2}-\frac{2}{n}\big)=
  \frac{1}{8G\mu}\big(\frac{n}{2}-\frac{2}{n}\big),
\end{eqnarray}
where we have used $c_L$ and $c_M$ central charges (\ref{cLM-GMMG}) into the right-hand sides of the equations in (\ref{cLM-GMMG2}).

\subsection{Single interval at zero temperature}\label{sec:4.1}
We consider a holographic $GCFT_2$ at zero temperature at the boundary of the space. A single interval is considered in the vacuum state of the field theory. The geometry of the bulk in context of the GMMG is a pure Minkowski spacetime. We use (\ref{eta.Min}) for the CS-contribution into the two-point correlator. The bulk extremal curve (\ref{EE02}) can be utilized into (\ref{EQ03}). Using the replica limit (\ref{EN02}) and $\chi_{n/2}$ and $\Delta_{n/2}$ (\ref{cLM-GMMG2}), we find the entanglement negativity for a single interval at zero temperature as follows
\begin{eqnarray}\label{EN1-zero}
  \mathcal{E} &=& \frac{c_L}{4}\log\big(\frac{2}{\epsilon}\sin\frac{\phi_{12}}{2}\big)+\frac{c_M}{8}\big |\frac{u_{12}}{\tan\frac{\phi_{12}}{2}} \big | \nonumber\\
  &=& \frac{3}{4G\mu}\log\big(\frac{2}{\epsilon}\sin\frac{\phi_{12}}{2}\big)-\frac{3}{8G}(\sigma+\frac{\alpha H}{\mu}+\frac{F}{m^2})\big |\frac{u_{12}}{\tan\frac{\phi_{12}}{2}} \big |
\end{eqnarray}
where we have used (\ref{cLM-GMMG2}) into the 2nd line of the entanglement negativity (\ref{EN1-zero}). Using the following transformations between the planar coordinates $(x,t)$ and the cylinder coordinates $(u,\phi)$
\begin{equation}\label{pl.Cy1}
  t=e^{i\phi},~~~x=iue^{i\phi},
\end{equation}
we can find the following relations
\begin{eqnarray}\label{pl.Cy2}
  t_{12} &=& 2\sin\frac{\phi_{12}}{2}\nonumber \\
  \frac{x_{12}}{t_{12}} &=& \frac{1}{2}\big |\frac{u_{12}}{\tan\frac{\phi_{12}}{2}} \big |.
\end{eqnarray}

Substituting relations (\ref{pl.Cy2}) into (\ref{EN1-zero}), we find the entanglement negativity in the plane coordinates as follows
\begin{eqnarray}\label{EN2-zero}
  \mathcal{E} &=& \frac{c_L}{4}\log\big(\frac{t_{12}}{\epsilon}\big)+\frac{c_M}{4}\frac{x_{12}}{t_{12}} \nonumber\\
  &=& \frac{3}{4G\mu}\log\big(\frac{t_{12}}{\epsilon}\big)-\frac{3}{2G}(\sigma+\frac{\alpha H}{\mu}+\frac{F}{m^2})\frac{x_{12}}{t_{12}}
\end{eqnarray}

Given that the central charges of the GMMG are different from the charges of the TMG, the result for the flat-TMG in \cite{Basu} can be extended to the GMMG in (\ref{EN2-zero}). The result (\ref{EN2-zero}) corresponds to the universal part of the result for the single interval configuration at zero temperature in \cite{Malv}.

To find the holographic entanglement entropy for a single interval at zero temperature, we can repeat the steps to find (\ref{EE-T01}). The two-point twist correlator is the same as (\ref{Q04}), only the scaling dimensions are different in GMMG in comparison with the TMG. Substituting the central charges $c_L$ and $c_M$ (\ref{cLM-GMMG2}) into the holographic entanglement entropy (\ref{EE-T01}), we find the following result for a zero-temperature single interval configuration
\begin{eqnarray}\label{EE1-zero}
  S_A &=& \frac{c_M}{12}S_{on-shell}^{EH}+\frac{c_L}{12}S_{on-shell}^{CS}\nonumber \\
   &=& \frac{c_M}{12}\big |\frac{u_{12}}{\tan\frac{\phi_{12}}{2}} \big |+\frac{c_L}{6}\log\big(\frac{2}{\epsilon}\sin\frac{\phi_{12}}{2}\big) \nonumber\\
   &=& -\frac{1}{4G}(\sigma+\frac{\alpha H}{\mu}+\frac{F}{m^2})\big |\frac{u_{12}}{\tan\frac{\phi_{12}}{2}} \big|+\frac{1}{2\mu G}\log\big(\frac{2}{\epsilon}\sin\frac{\phi_{12}}{2}\big)
\end{eqnarray}
where we have used (\ref{L01}) and (\ref{eta.Min}) in 2nd line. By comparing the holographic entanglement negativity (\ref{EN1-zero}) and the holographic entanglement entropy (\ref{EE1-zero}), we find the following relation between these results
\begin{equation}\label{E&N1}
  \mathcal{E}=\frac{3}{2}S_A.
\end{equation}

This result is consistent with the universal part of the result in \cite{Malv}.

\subsection{Single interval at a finite-sized system}\label{sec:4.2}
In this subsection we find the holographic entanglement negativity for the bipartite state of a single interval in a $GCFT_2$ with finite-sized system with periodic boundary conditions. The bulk geometry is a global Minkowski spacetime in the null infinity of the GMMG. Instead of the metric (\ref{glob.met}), we have the following one \cite{Jiang}
\begin{equation}\label{met.orb}
  ds^2=dr^2-(\frac{2\pi}{l_{\phi}})^2 du^2+r^2d\phi^2.
\end{equation}
where the boundary periodic condition is as follows \cite{Jiang}
\begin{equation}\label{orb.b}
  (u,\phi)\sim(u,\phi+l_{\phi}).
\end{equation}

To find the entanglement negativity and the entanglement entropy, we need $\Delta\eta_{12}$ and $L^{extr}_{tot}$ that we can use our previous result in \cite{Setare3}. For a field theory with following identification
\begin{equation}\label{thermalid}
  (u,\phi)\sim (u+i\beta_u,\phi-i\beta_{\phi})
\end{equation}
we have the entanglement entropy for a single interval as follows
\begin{equation}\label{EE.bef}
  S_A=\frac{c_L}{6}\log \big(\frac{\beta_{\phi}}{\pi\epsilon_{\phi}}\sinh \frac{\pi l_{\phi}}{\beta_{\phi}} \big)
  +\frac{c_M}{6}\frac{1}{\beta_{\phi}}\big(\pi(l_u+\frac{\beta_{u}}{\beta_{\phi}}l_\phi)\coth\frac{\pi l_{\phi}}{\beta_{\phi}}-\beta_u\big)-\frac{c_M}{6}\frac{\epsilon_u}{\epsilon_{\phi}}.
\end{equation}

By comparing (\ref{orb.b}) and (\ref{thermalid}), we find the following identifications to turn the entanglement entropy (\ref{EE.bef}) for a finite-sized system,
\begin{eqnarray}\label{idD}
  \beta_u &=& 0,~~~\beta_{\phi}=iL_{\phi},\nonumber \\
  l_{\phi} &=& \phi_{12} \nonumber\\
  l_u &=& u_{12}
\end{eqnarray}

Substituting (\ref{idD}) into (\ref{EE.bef}), we find the entanglement entropy for a single interval in a $GCFT_2$ with finite-sized system as follows
 \begin{equation}\label{EE1.fin}
  S_A=\frac{c_L}{6}\log \big(\frac{L_{\phi}}{\pi\epsilon}\sin \frac{\pi \phi_{12}}{L_{\phi}} \big)
  +\frac{c_M}{6}\frac{\pi u_{12}}{L_{\phi}}\cot\frac{\pi \phi_{12}}{L_{\phi}}.
\end{equation}

By comparing (\ref{EE1.fin}) and (\ref{idD}), we find the following results for the extremal curve length  $L^{extr}_{tot}$ and the extremized boost $\Delta\eta_{12}$ as follows
\begin{equation}\label{L.fin}
  L^{extr}_{tot}=\frac{2\pi u_{12}}{L_{\phi}}\cot\frac{\pi \phi_{12}}{L_{\phi}} \\
\end{equation}
 and
 \begin{equation}\label{eta.fin}
  \Delta\eta_{12}= 2\log \big(\frac{L_{\phi}}{\pi\epsilon}\sin \frac{\pi \phi_{12}}{L_{\phi}} \big)
\end{equation}

 The bulk extremal curve (\ref{L.fin}) and the spinning contribution (\ref{eta.fin}) can be utilized into the two-point correlator (\ref{EQ03}). Utilizing the replica technique (\ref{EN02}), we find the entanglement negativity as follows
\begin{eqnarray}\label{EN1.fin}
  \mathcal{E} &=& \frac{c_M}{8}L^{extr}_{tot}+\frac{c_L}{8}\Delta\eta_{12} \nonumber \\
   &=& \frac{c_M}{4}\frac{\pi u_{12}}{L_{\phi}}\cot\frac{\pi \phi_{12}}{L_{\phi}}
   +\frac{c_L}{4}\log \big(\frac{L_{\phi}}{\pi\epsilon}\sin \frac{\pi \phi_{12}}{L_{\phi}} \big)
\end{eqnarray}

This result is the same as the result in flat-TMG in \cite{Basu}, where the only difference is in the central charges between GMMG and TMG. We can utilize the central charges (\ref{cLM-GMMG}) into (\ref{EN1.fin}) that the following relation for the entanglement negativity can be found
\begin{equation}\label{EN2.fin}
  \mathcal{E} = -\frac{3}{4G}(\sigma+\frac{\alpha H}{\mu}+\frac{F}{m^2})\frac{\pi u_{12}}{L_{\phi}}\cot\frac{\pi \phi_{12}}{L_{\phi}}
   +\frac{3}{4G\mu}\log \big(\frac{L_{\phi}}{\pi\epsilon}\sin \frac{\pi \phi_{12}}{L_{\phi}} \big)
\end{equation}

Substituting the central charges (\ref{cLM-GMMG}) into the entanglement entropy (\ref{EE1.fin}), we find the following result
\begin{equation}\label{EE2.fin}
 S_A = -\frac{1}{2G}(\sigma+\frac{\alpha H}{\mu}+\frac{F}{m^2})\frac{\pi u_{12}}{L_{\phi}}\cot\frac{\pi \phi_{12}}{L_{\phi}}
   +\frac{1}{2G\mu}\log \big(\frac{L_{\phi}}{\pi\epsilon}\sin \frac{\pi \phi_{12}}{L_{\phi}} \big).
\end{equation}

By comparing (\ref{EN2.fin}) and (\ref{EE2.fin}), the following relation between the entanglement entropy and the entanglement negativity for a single interval in a finite-sized system can be found
\begin{equation}\label{E&N2}
  \mathcal{E}=\frac{3}{2}S_A.
\end{equation}

This result is the same as the universal part of the result of \cite{Malv} that has been obtained by another approach there.

\subsection{Single interval at a finite temperature}\label{sec:4.3}
For a single interval $A$ at a finite temperature, the $GCFT_2$ can be considered as a thermal field theory on a cylinder with complex circumference $\beta$. A naive two-point twist correlator as (\ref{EQ03}) to compute the entanglement negativity leads to an incorrect result \cite{Malv}. For this reason, two other finite intervals $B_1$ and $B_2$ adjacent to the interval $A$ are necessary to be considered. The bulk spacetime is described by GMMG in Flat Space Cosmology (FSC) geometries. To find the entanglement negativity, we compute the following four-point twist correlator instead of the two-point correlator (\ref{EQ03})
\begin{eqnarray}\label{4p-temp01}
  \langle\Phi_n(x_1,t_1)\Phi^2_{-n}(x_2,t_2)\Phi_n^2(x_3,t_3)\Phi_{-n}(x_4,t_4)\big\rangle &=&
  \frac{t_{13}^{\Delta_n+\Delta_n^{(2)}}t_{24}^{\Delta_n+\Delta_n^{(2)}}}{t_{12}^{\Delta_n+\Delta_n^{(2)}}
  t_{23}^{\Delta_n+\Delta_n^{(2)}}t_{34}^{\Delta_n+\Delta_n^{(2)}}t_{14}^{2\Delta_n}}\nonumber \\
   && \times \exp\bigg[\frac{x_{13}}{t_{13}}(\chi_n+\chi_n^{(2)})+\frac{x_{24}}{t_{24}}(\chi_n+\chi_n^{(2)})
   \nonumber \\
   && -\frac{x_{12}}{t_{12}}(\chi_n+\chi_n^{(2)}) -\frac{x_{23}}{t_{23}}(2\chi_n)\nonumber\\
   && -\frac{x_{34}}{t_{34}}(\chi_n+\chi_n^{(2)})-\frac{x_{14}}{t_{14}}(2\chi_n) \bigg]\mathcal{G}(t,\frac{x}{t})\nonumber\\
   &,&
\end{eqnarray}
where we have just written the universal part of the four-point correlator. To write the right-hand side of (\ref{4p-temp01}), we have used the general form of the four-point twist field correlator in \cite{Malv}. Substituting the following identities
\begin{equation}\label{ids2}
  (\frac{t_{13}t_{24}}{t_{12}t_{34}})^{\Delta_n}\equiv t^{-\Delta_n}
\end{equation}
and
\begin{equation}\label{ids3}
  \chi_n(\frac{x_{12}}{t_{12}}+\frac{x_{34}}{t_{34}}-\frac{x_{13}}{t_{13}}-\frac{x_{24}}{t_{24}})\equiv \chi_n \frac{x}{t}
\end{equation}
and inserting these identities into the $\mathcal{G}(t,\frac{x}{2})$, we can find the four-point correlator
as follows
\begin{eqnarray}\label{4p-temp02}
  \langle\Phi_n(x_1,t_1)\Phi^2_{-n}(x_2,t_2)\Phi_n^2(x_3,t_3)\Phi_{-n}(x_4,t_4)\big\rangle &=& \exp\bigg[-\chi_nL_{14}^{extr}-\chi_{n/2}\big(2L^{extr}_{23}+L^{extr}_{12}+L^{extr}_{34}\nonumber \\
   && -L^{extr}_{13}-L^{extr}_{24}\big)-\Delta_n\Delta\eta_{14}-\Delta_{n/2}\big(2\Delta\eta_{23}\nonumber \\
   && +\Delta\eta_{12}+\Delta\eta_{34}-\Delta\eta_{13}-\Delta\eta_{14} \big)\bigg]
\end{eqnarray}
where we have used the following relations
\begin{equation}\label{ids4}
  L_{ij}^{extr}=2\frac{x_{ij}}{t_{ij}}
\end{equation}
and
\begin{equation}\label{ids5}
  \Delta\eta_{ij}=2\log\frac{t_{ij}}{\epsilon},
\end{equation}
for the extremal curve length and the extremized boost, respectively.

Substituting (\ref{4p-temp02}) into the replica technique (\ref{EN02}), we can find the entanglement negativity as follows
\begin{eqnarray}\label{EN1-temp}
  \mathcal{E} &=& \frac{c_M}{16}\big(2L_A+L_{B_1}+L_{B_2}-L_{A\cup B_1}-L_{A\cup B_2}\big) \\
   &+& \frac{c_L}{16}\big(2\Delta\eta_A+\Delta\eta_{B_1}+\Delta\eta_{B_2}-\Delta\eta_{A\cup B_1}-\Delta\eta_{A\cup B_2} \big)
\end{eqnarray}
where we have used the following definitions \cite{Basu}
\begin{eqnarray}
  L_{12}^{extr} &=& L_{B_1},~~L_{23}^{extr}=L_A,~~L_{34}^{extr}=L_{B_2}, \\
   L_{13}^{extr} &=& L_{A\cup B_1},~~L_{24}^{extr}=L_{A\cup B_2},~~L_{14}^{extr}=L_{A\cup B}.
\end{eqnarray}

Substituting the central charges (\ref{cLM-GMMG}) into (\ref{EN1-temp}), we find the following result for the entanglement negativity
\begin{equation}\label{EN2-temp}
  \mathcal{E}=\frac{3}{16G}\big(2\mathcal{L}_A+\mathcal{L}_{B_1}+\mathcal{L}_{B_2}-\mathcal{L}_{A\cup B_1}
  -\mathcal{L}_{A\cup B_2}  \big)
\end{equation}
where
\begin{equation}\label{LX}
  \mathcal{L}_X=-(\sigma+\frac{\alpha H}{\mu}+\frac{F}{m^2})L_X+\frac{1}{\mu}\Delta\eta_X.
\end{equation}

To find the exact form of the universal part of the entanglement negativity, one can write the right-hand side of (\ref{4p-temp01}) as follows
\begin{eqnarray}\label{4p-temp03}
  \langle\Phi_n(x_1,t_1)\Phi^2_{-n}(x_2,t_2)\Phi_n^2(x_3,t_3)\Phi_{-n}(x_4,t_4)\big\rangle &=&
  t^{-\Delta_n}t^{-\Delta^{(2)}_n}t^{-2\Delta_n}_{14}t^{-2\Delta^{(2)}}_{23}\nonumber \\
   &\times& \exp\big[-\chi_n^{(2)}\frac{x}{t}-\chi_n\frac{x}{t}\nonumber\\
   &-& 2\chi_n^{(2)}\frac{x_{23}}{t_{23}}-2\chi_n\frac{x_{14}}{t_{14}}\big].
\end{eqnarray}

In the bipartite limit ($L\to \infty$), we have the following relations for $x$ and $\frac{x}{t}$ as follows \cite{Malv}
\begin{equation}\label{b-lim}
  \lim_{L\to \infty}t=\exp(-\frac{2\pi}{\beta}\phi_{23}),~~~\lim_{L\to \infty}\frac{x}{t}=-\frac{2\pi }{\beta}u_{23}.
\end{equation}

Utilizing the entanglement entropy (\ref{EE.bef}) that we have computed in our previous work \cite{Setare3} for a finite temperature field theory, we can find the following result
\begin{eqnarray}\label{t01}
  t_{ij} &=& \frac{\beta}{\pi}\sinh\big(\frac{\pi}{\beta}\phi_{ij}\big) \\
  \frac{x_{ij}}{t_{ij}} &=& \frac{\pi}{\beta}u_{ij}\coth\big(\frac{\pi}{\beta}\phi_{ij}\big) \label{xt01}
\end{eqnarray}
where we have used the following identifications into (\ref{EE.bef})
\begin{eqnarray}\label{idD2}
  \beta_u &=& 0,~~~\beta_{\phi}=\beta,\nonumber \\
  l_{\phi} &=& \phi_{12}, \nonumber\\
  l_u &=& u_{12}.
\end{eqnarray}

Substituting (\ref{t01}) and (\ref{b-lim}) into (\ref{4p-temp03}) and using the replica limit (\ref{EN02}), we find the entanglement negativity for a finite temperature $GCFT_2$ as follows
\begin{eqnarray}\label{EN3-temp}
  \mathcal{E} &=& \frac{c_L}{4}\log\big(\frac{\beta}{\pi \epsilon}\sinh\big(\frac{\pi}{\beta}\phi_{23}\big)  \big)- \frac{c_L}{4}\frac{\pi}{\beta}\phi_{23} \nonumber\\
   &+& \frac{c_M}{4}\frac{\pi}{\beta}u_{23}\coth\big(\frac{\pi}{\beta}\phi_{23}   \big)-\frac{c_M}{4}\frac{\pi}{\beta}u_{23}
\end{eqnarray}

This is same as the result in \cite{Basu} where the only difference is in the central charges (\ref{cLM-GMMG}). The result (\ref{EN3-temp}) is compatible with the universal part of the result in \cite{Malv}. To compute the entanglement entropy using the replica technique (\ref{EE-T01}), we use (\ref{t01}) and (\ref{xt01}) to find the extremal curve length $L^{extr}_{FSC}$ and the extremized boost $\Delta\eta_{ij}$
\begin{equation}\label{L-temp}
  L^{extr}_{FSC}= 2\frac{\pi}{\beta}u_{23}\coth\big(\frac{\pi}{\beta}\phi_{23}\big)
\end{equation}
and
\begin{equation}\label{CS-temp}
  S^{CS}_{FSC}=2\log\bigg(\frac{\beta}{\pi\epsilon}\sinh\big(\frac{\pi}{\beta}\phi_{23}\big) \bigg).
\end{equation}

 Substituting (\ref{L-temp}) and (\ref{CS-temp}) into two-point correlator (\ref{Q04}) and utilizing the replica technique (\ref{EE-T01}), we find the holographic entanglement entropy as follows
\begin{eqnarray}\label{EE02-temp}
  S_A &=& \frac{c_M}{6} \frac{\pi}{\beta}u_{23}\coth\big(\frac{\pi}{\beta}\phi_{23}   \big)
  + \frac{c_L}{6}\log\big(\frac{\beta}{\pi \epsilon}\sinh\big(\frac{\pi}{\beta}\phi_{23}\big)\big) \nonumber  \\
   &=& -\frac{1}{2G}(\sigma+\frac{\alpha H}{\mu}+\frac{F}{m^2})\frac{\pi}{\beta}u_{23}\coth\big(\frac{\pi}{\beta}\phi_{23}   \big)\nonumber \\
   &+& \frac{1}{2G\mu}\log\big(\frac{\beta}{\pi \epsilon}\sinh\big(\frac{\pi}{\beta}\phi_{23}\big)\big).
\end{eqnarray}

By comparing (\ref{EN3-temp})and (\ref{EE02-temp}), one can find the following result between the holographic entanglement entropy and the holographic entanglement negativity
\begin{equation}\label{3S}
  \mathcal{E}=\frac{3}{2}\big(S_A-S_{th}  \big)
\end{equation}
where $S_{th}$ is the thermal entropy and has the following form
\begin{equation}\label{ther.E}
  S_{th}= \frac{c_L}{4}\frac{\pi}{\beta}\phi_{23}+\frac{c_M}{4}\frac{\pi}{\beta}u_{23}.
\end{equation}
where can be calculated by the Cardy formula \cite{car1,car2}. The Cardy formula for the thermal entropy can be considered as follows
\begin{equation}\label{Card01}
  S_{th}= -\frac{\pi^2}{3}\big(c_L\frac{b}{a}+c_M\frac{\bar{a}b-a\bar{b}}{a^2}  \big),
\end{equation}
where there is the following identification between the thermal circle and the spatial circle
\begin{equation}\label{th.sp01}
  (\tilde{u},\tilde{\phi})\sim(\tilde{u}+i\bar{a},\tilde{\phi}-ia)
  \sim(\tilde{u}+2\pi\bar{b},\tilde{\phi}-2\pi b),
\end{equation}
where $(a,\bar{a})$ parameterizes the thermal circle and $(b,\bar{b})$ parameterizes the spatial circle. The formula (\ref{Card01}) is the general form of the thermal entropy. We can use this formula in the configuration of a single interval at finite temperature $GCFT_2$, utilizing the following identification
\begin{equation}\label{th.sp02}
  (u,\phi)\sim(u,\phi-i\beta)
  \sim(u+u_{23},\phi+\phi_{23}).
\end{equation}

By comparing (\ref{th.sp01}) and (\ref{th.sp02}), we find the special parameters of the circles. Substituting the parameters into (\ref{Card01}), the result (\ref{ther.E}) for the thermal entropy can be established. The result (\ref{3S}) in the FSC background of the GMMG and for the finite temperature $GCFT_2$ is the same as the result in \cite{Malv} in the $GCFT_2$ with a bipartite system of a single interval at finite temperature.

\section{Entanglement negativity and entanglement entropy for adjacent intervals  }\label{sec:5}
In this section, we consider the $GCFT_2$ on the boundary with two adjacent intervals $A_1$ and $A_2$. To compute the entanglement negativity, the GMMG can be considered as the bulk gravity theory with asymptotically flat space geometry. As we have two adjacent intervals, the four-point twist correlator (\ref{Q01}) can be used to find the entanglement negativity. Since these intervals $A_1$ and $A_2$ are not far apart, the correlator (\ref{Q01}) can be turned into the following form
\begin{equation}\label{Q-ad01}
  \mathrm{Tr}( \rho_A^{T_2})^{n}=\langle\Phi_n(x_1,t_1)\Phi^2_{-n}(x_2,t_2)\Phi_{n}(x_3,t_3) \rangle,
\end{equation}
where we have utilized the following limit into (\ref{Q01})
\begin{equation}\label{lim-2ad}
  (x_2,t_2)\to (x_3,t_3).
\end{equation}

By applying the limit (\ref{lim-2ad}) into the four-point correlator, we can find the three-point correlator (\ref{Q-ad01}) as follows
\begin{eqnarray}\label{Q-ad02}
  \langle\Phi_n(x_1,t_1)\Phi^2_{-n}(x_2,t_2)\Phi_{n}(x_3,t_3)\big\rangle &=& \exp\bigg[-\chi_nL_{13}^{extr}-\chi_{n/2}\big(L^{extr}_{12}+L^{extr}_{23}-L^{extr}_{13}\big)\nonumber \\
   &-& \Delta_n\Delta\eta_{13}-\Delta_{n/2}\big(\Delta\eta_{12}+\Delta\eta_{23}-\Delta\eta_{13} \big)\bigg]
\end{eqnarray}

Utilizing the central charges $c_L$ and $c_M$ (\ref{cLM-GMMG}) into (\ref{Q-ad02}) and by substituting (\ref{Q-ad02}) into the replica limit (\ref{EN02}), the holographic entanglement negativity can be found as follows
\begin{equation}\label{EN01-adj}
  \mathcal{E}=\frac{3}{16G}\bigg[-(\sigma+\frac{\alpha H}{\mu}+\frac{F}{m^2})\big(L^{extr}_{12}+L^{extr}_{23}-L^{extr}_{13}\big)
  +\frac{1}{\mu}\big(\Delta\eta_{12}+\Delta\eta_{23}-\Delta\eta_{13}   \big)  \bigg]
\end{equation}

This equation can be expressed as follows
\begin{equation}\label{EN02-adj}
  \mathcal{E}=\frac{3}{16G}\bigg[\mathcal{L}_{A_1}+\mathcal{L}_{A_2}-\mathcal{L}_{A_1\cup A_2} \bigg]
\end{equation}
where we have defined $\mathcal{L}_{X}$ as follows
\begin{equation}\label{LX-adj}
  \mathcal{L}_{X}=-(\sigma+\frac{\alpha H}{\mu}+\frac{F}{m^2})L_X+\frac{1}{\mu}\Delta\eta_X.
\end{equation}

This result is the same as the entanglement negativity of two adjacent intervals in \cite{Basu}, where the only difference is the scaling dimensions of the twist fields.

\subsection{Adjacent intervals at zero temperature}\label{sec:5.1}
To consider the $GCFT_2$ at zero temperature, the two adjacent intervals should be in the vacuum state of the theory on the boundary. We have computed the extremal curve length $L_{tot}^{extr}$ and the spinning contribution $\Delta\eta_{ij}$ in the FSC bulk geometry in (\ref{ids4}) and (\ref{ids5}), respectively. Substituting $L_{tot}^{extr}$ and $\Delta\eta_{ij}$ into (\ref{EN01-adj}),  we find the holographic entanglement negativity as follows
\begin{equation}\label{EN01-adj0}
  \mathcal{E}=-\frac{3}{8G}(\sigma+\frac{\alpha H}{\mu}+\frac{F}{m^2})\big(\frac{x_{12}}{t_{12}}+\frac{x_{23}}{t_{23}}-\frac{x_{13}}{t_{13}}\big)
    +\frac{3}{8G}\log\frac{t_{12}t_{23}}{\epsilon t_{13}} .
\end{equation}

We can write (\ref{EN01-adj0}) with the central charges (\ref{cLM-GMMG}) as follows
\begin{equation}\label{EN02-adj0}
  \mathcal{E}=\frac{c_M}{8}\big(\frac{x_{12}}{t_{12}}+\frac{x_{23}}{t_{23}}-\frac{x_{13}}{t_{13}}\big)
  +\frac{c_L}{8}\log\frac{t_{12}t_{23}}{\epsilon (t_{12}+t_{23})}.
\end{equation}

The holographic entanglement negativity (\ref{EN02-adj0}) is the same as the result for flat-TMG in \cite{Basu}, but with different central charges. To compute the holographic entanglement entropy, we utilize the general form of the entropy in (\ref{EE-T01}) as follows
\begin{equation}\label{EE-gen}
  S_A=\frac{c_M}{12}L_{ij}^{extr}+\frac{c_L}{12}\Delta\eta_{ij},
\end{equation}

Utilizing (\ref{ids4}) and (\ref{ids5}) into (\ref{EE-gen}), we find the entanglement entropy of a single interval $[(x_i,t_i),(x_j,t_j)]$ in FSC bulk geometry as follows
\begin{equation}\label{EE-gen2}
  S_{ij}=\frac{c_M}{6}\frac{x_{ij}}{t_{ij}}+\frac{c_L}{6}\log\frac{t_{ij}}{\epsilon}.
\end{equation}

By comparing the entanglement negativity (\ref{EN02-adj0}) and the entanglement entropy (\ref{EE-gen2}), we can easily find the following relation between them
\begin{equation}\label{EE-EN3}
  \mathcal{E}=\frac{3}{4}(S_{12}+S_{23}-S_{13}),
\end{equation}
where $S_{12}$, $S_{23}$ and $S_{13}$ can be found by (\ref{EE-gen2}). The relation between the entanglement entropy and the entanglement negativity in (\ref{EE-EN3}) is same as the result between the two types of the entropies in \cite{Malv}.

\subsection{Adjacent intervals at a finite-sized system}\label{sec:5.2}
We find the holographic entanglement negativity for a configuration of two adjacent intervals in a $GCFT_2$ with a finite size. The $GCFT_2$ is the boundary side theory of the duality that can be considered on an infinite cylinder with $L_{\phi}$ circumference \cite{Basu}. The gravity model in the bulk is the GMMG model in FSC geometry.  We utilize (\ref{L.fin}) and (\ref{eta.fin}) for extremal curve length $L_{ij}^{extr}$ and the extremized boost $\Delta\eta_{ij}$, respectively. Comparing (\ref{ids4}) and (\ref{ids5}) with (\ref{L.fin}) and (\ref{eta.fin}), we find the following result
\begin{equation}\label{xt02}
  \frac{x_{ij}}{t_{ij}}=\frac{\pi u_{ij}}{L_{\phi}}\cot\frac{\pi \phi_{ij}}{L_{\phi}} \\
\end{equation}
 and
 \begin{equation}\label{t02}
  t_{ij}=  \frac{L_{\phi}}{\pi}\sin \frac{\pi \phi_{ij}}{L_{\phi}}.
\end{equation}
Substituting (\ref{xt02}) and (\ref{t02}) into the entanglement negativity relation (\ref{EN02-adj0})
\begin{eqnarray}\label{EN01-adjFS}
  \mathcal{E} &=& \frac{c_L}{8}\log\bigg(\frac{L_{\phi}}{\pi \epsilon}\frac{\sin \frac{\pi \phi_{12}}{L_{\phi}}\sin \frac{\pi \phi_{23}}{L_{\phi}}}{\sin \frac{\pi \phi_{13}}{L_{\phi}}} \bigg)\nonumber \\
   &+& \frac{c_M}{8}\bigg(\frac{\pi u_{12}}{L_{\phi}}\cot\frac{\pi \phi_{12}}{L_{\phi}}
   +\frac{\pi u_{23}}{L_{\phi}}\cot\frac{\pi \phi_{23}}{L_{\phi}}-\frac{\pi u_{13}}{L_{\phi}}\cot\frac{\pi \phi_{13}}{L_{\phi}}  \bigg)
\end{eqnarray}

This is same as the result in \cite{Basu}. The result (\ref{EN01-adjFS}) for asymptotically flat GMMG is different with the TMG result in flat TMG. The only difference is due to the different central charges for these massive gravity models. Substituting (\ref{t02}) and (\ref{xt02}) into (\ref{EE-gen2}), we can reproduce (\ref{EE-EN3}) for the finite-sized system. This result is consistent with the result in \cite{Malv}.

\subsection{Adjacent intervals at a finite temperature}\label{sec:5.3}
In this subsection, we focus on the configuration of two adjacent intervals in a finite temperature $GCFT_2$. As the previous section, we have the theory on an infinite cylinder but with complex circumference $\beta$. The bulk geometry at null infinity is FSC geometry that is a solution of GMMG model. To compute the following holographic entanglement negativity, we utilize relations (\ref{t01}) and (\ref{xt01}) into (\ref{EN02-adj0})
\begin{eqnarray}\label{EN01-adjTEMP}
  \mathcal{E} &=& \frac{c_L}{8}\log\bigg(\frac{\beta}{\pi \epsilon}\frac{\sinh \frac{\pi \phi_{12}}{\beta}\sinh \frac{\pi \phi_{23}}{\beta}}{\sinh \frac{\pi \phi_{13}}{\beta}} \bigg)\nonumber \\
   &+& \frac{c_M}{8}\bigg(\frac{\pi u_{12}}{\beta}\coth\frac{\pi \phi_{12}}{\beta}
   +\frac{\pi u_{23}}{\beta}\coth\frac{\pi \phi_{23}}{\beta}-\frac{\pi u_{13}}{\beta}\coth\frac{\pi \phi_{13}}{\beta}  \bigg).
\end{eqnarray}

Except for the difference in the central charges, there is no difference between the result (\ref{EN01-adjTEMP}) in the GMMG and the entanglement negativity of two adjacent intervals in a finite temperature case in \cite{Basu}. Substituting relations (\ref{L-temp}) and (\ref{CS-temp}) into the holographic entanglement entropy (\ref{EE-gen2}) for a single interval, we find the following result
\begin{equation}\label{EE-sing01}
  S_{ij}=\frac{c_M}{6} \frac{\pi}{\beta}u_{ij}\coth\big(\frac{\pi}{\beta}\phi_{ij}   \big)
  + \frac{c_L}{6}\log\big(\frac{\beta}{\pi \epsilon}\sinh\big(\frac{\pi}{\beta}\phi_{ij}\big)\big).
\end{equation}

By comparing (\ref{EN01-adjTEMP}) and (\ref{EE-sing01}), the result (\ref{EE-EN3}) can be reproduced, that is the same as the result in \cite{Malv}. We have had this result in all three cases of the two adjacent intervals configurations.

\section{Entanglement negativity and entanglement entropy for disjoint intervals  }\label{sec:6}
In this section, we compute the holographic entanglement negativity and the holographic entanglement entropy for the state of two disjoint intervals in a $GCFT_2$. In the bulk side of the duality, we have the GMMG in pure Minkowski spacetime. As we have two disjoint intervals, the four-point twist correlator (\ref{Q01}) can be used to find the entanglement negativity. The four point twist correlator for two disjoint intervals $A_1=[(x_1,t_1),(x_2,t_2)]$ and $A_2=[(x_3,t_3),(x_4,t_4)]$ can be considered as follows \cite{Basu}
\begin{eqnarray}\label{4p-2d01}
  \langle\Phi_n(x_1,t_1)\Phi_{-n}(x_2,t_2)\Phi_n(x_3,t_3)\Phi_{-n}(x_4,t_4)\big\rangle &=&
 \exp\bigg[ \frac{c_M}{16}\big(L^{extr}_{13}+L^{extr}_{24}-L^{extr}_{14}-L^{extr}_{23}     \big)\nonumber\\
 &+& \frac{c_L}{16}\big(\Delta\eta_{13}+\Delta\eta_{24}-\Delta\eta_{14}-\Delta\eta_{23}     \big)\bigg].
\end{eqnarray}

Substituting (\ref{4p-2d01}) into the replica limit (\ref{EN02}), we find the entanglement negativity as follows
\begin{equation}\label{EN01-2i0}
  \mathcal{E} =
  \frac{3}{16G}\big(\mathcal{L}^{extr}_{13}+\mathcal{L}^{extr}_{24}-\mathcal{L}^{extr}_{14}-\mathcal{L}^{extr}_{23}     \big),
\end{equation}
where
\begin{equation}\label{L001}
  \mathcal{L}^{extr}_{X}=-(\sigma+\frac{\alpha H}{\mu}+\frac{F}{m^2})L^{extr}_{X}+\frac{1}{\mu}\Delta\eta_X.
\end{equation}

We can use this result to compute the entanglement negativity in different configurations involving two disjoint intervals.

\subsection{Disjoint intervals at zero temperature}\label{sec:6.1}
To consider the two disjoint intervals configurations, the $GCFT_2$ should be in the ground state that is dual to the GMMG in pure Minkowski spacetime. By comparing (\ref{EN1-zero}) and (\ref{pl.Cy2}), we use the extremal curve and the spinning contribution in the pure Minkowski spapcetime as follows
\begin{equation}\label{L.et.2i01}
  L_{ij}^{extr}=2\frac{x_{ij}}{t_{ij}},~~~\Delta\eta_{ij}=2\log(\frac{t_{ij}}{\epsilon}).
\end{equation}

Substituting the extremal curve length $L_{ij}^{extr}$ and the extremal boost $\Delta\eta_{ij}$ (\ref{L.et.2i01}) into the entanglement negativity (\ref{EN01-2i0}), we find the entanglement negativity for two disjoint intervals at zero temperature as follows
\begin{equation}\label{EN02-2i0}
  \mathcal{E}=-\frac{3}{8G}(\sigma+\frac{\alpha H}{\mu}+\frac{F}{m^2})\big(\frac{x_{13}}{t_{13}}+
  \frac{x_{24}}{t_{24}}-\frac{x_{14}}{t_{14}}-\frac{x_{23}}{t_{23}}   \big)+\frac{3}{8G\mu}
  \log\big(\frac{t_{13}t_{24}}{t_{14}t_{23}}   \big)
\end{equation}
where we can write this result as following
\begin{equation}\label{EN03-2i0}
  \mathcal{E}=\frac{c_M}{8}\big(\frac{x_{13}}{t_{13}}+
  \frac{x_{24}}{t_{24}}-\frac{x_{14}}{t_{14}}-\frac{x_{23}}{t_{23}}   \big)+\frac{c_L}{8}
  \log\big(\frac{t_{13}t_{24}}{t_{14}t_{23}}   \big).
\end{equation}

This result is same as the result for two disjoint intervals configurations at zero temperature holographic $GCFT_2$ in \cite{Basu}.

\subsection{Disjoint intervals at a finite-sized system}\label{sec:6.2}
In this subsection, the two disjoint intervals are considered in a finite-sized system. The boundary theory is on a cylinder with circumference $L_{\phi}$. The gravity in the bulk side of the theory is described by FSC geometry in the GMMG. Substituting (\ref{xt02}) and (\ref{t02}) into (\ref{ids4}) and (\ref{ids5}) respectively, we find the extremal curve and the spinning contribution as follows
\begin{eqnarray}\label{L.et.2i001}
  L_{ij}^{extr} &=& \frac{2\pi}{L_{\phi}}u_{ij}\cot\big(\frac{\pi}{l_{\phi}}\phi_{ij}   \big), \nonumber\\
  \Delta\eta_{ij} &=& 2\log\big(\frac{L_{\phi}}{\pi}\sin\frac{\pi}{l_{\phi}}\phi_{ij}    \big).
\end{eqnarray}

Substituting (\ref{L.et.2i001}) into (\ref{EN01-2i0}), we find the entanglement negativity as follows
\begin{eqnarray}\label{EN03-2i0}
  \mathcal{E} &=& \frac{c_L}{8}\log\bigg(\frac{L_{\phi}}{\pi}\frac{\sin \frac{\pi \phi_{13}}{L_{\phi}}\sin \frac{\pi \phi_{24}}{L_{\phi}}}{\sin \frac{\pi \phi_{14}}{L_{\phi}}\sin \frac{\pi \phi_{23}}{L_{\phi}}} \bigg)\nonumber \\
   &+& \frac{c_M}{8}\bigg(\frac{\pi u_{13}}{L_{\phi}}\cot\frac{\pi \phi_{13}}{L_{\phi}}
   +\frac{\pi u_{24}}{L_{\phi}}\cot\frac{\pi \phi_{24}}{L_{\phi}}-\frac{\pi u_{14}}{L_{\phi}}\cot\frac{\pi \phi_{14}}{L_{\phi}}-\frac{\pi u_{23}}{L_{\phi}}\cot\frac{\pi \phi_{23}}{L_{\phi}}  \bigg).
\end{eqnarray}

\subsection{Disjoint intervals at a finite temperature}\label{sec:6.3}
In this subsection, we have a configuration state of two disjoint intervals in a finite temperature $GCFT_2$. The boundary is living on a cylinder with complex circumference $\beta$. The bulk side of the duality is described by the FSC geometry in the GMMG. To find the holographic entanglement entropy, we utilize the extremal curve (\ref{L-temp}) and the spinning contributions (\ref{CS-temp}) into (\ref{EN01-2i0}) as follows
\begin{eqnarray}\label{EN04-2i0}
  \mathcal{E} &=& \frac{c_L}{8}\log\bigg(\frac{\beta}{\pi}\frac{\sinh \frac{\pi \phi_{13}}{\beta}\sinh \frac{\pi \phi_{24}}{\beta}}{\sinh \frac{\pi \phi_{14}}{\beta}\sinh \frac{\pi \phi_{23}}{\beta}} \bigg)\nonumber \\
   &+& \frac{c_M}{8}\bigg(\frac{\pi u_{13}}{\beta}\coth\frac{\pi \phi_{13}}{\beta}
   +\frac{\pi u_{24}}{\beta}\coth\frac{\pi \phi_{24}}{\beta}-\frac{\pi u_{14}}{\beta}\coth\frac{\pi \phi_{14}}{\beta}\nonumber\\
   &-& \frac{\pi u_{23}}{\beta}\coth\frac{\pi \phi_{23}}{\beta}  \bigg).
\end{eqnarray}

\section{Conclusion}\label{sec:7}
In this paper, we consider a class of non-relativistic gauge/gravity dualities in $2+1$ dimension that could be interesting in the condensed matter physics, nuclear physics and other real physical systems. The symmetry, which governs the asymptotic behavior of the gravity and the vacuum behavior of the boundary field theory, has been considered by the Galilean conformal group which is a parametric contraction of conformal group.
 The Galilean conformal field theory ($GCFT_2$) \cite{repl3,repl4} concerns the two-dimensional boundary of the holography  which is dual with the generalized minimal massive gravity (GMMG)\cite{Setare1}. GMMG model that can be considered as the gravity that descibes the bulk geometry, would be an extension of MMG model \cite{mmg} by adding the CS-term and an extra term to pure Einstein gravity with a negative cosmological constant.
 The bulk geometry has been considered as an asymptotically flat spacetime which is a solution of GMMG model. The asymptotic symmetry
   of this background is described by the Galilean conformal algebra (GCA) in two dimensions \cite{Setare2} in which the appropriate central charges should be placed. The same algebra with the mentioned central charges will also apply to the $GCFT_2$. Using the GCA and the twist field weights in GCFT, correlation functions are constructed that play an important role in investigating the properties of quantum entanglement.

The calculation of quantum entanglement provides the context which is important for the study of various phenomena in a large area from the physics of condensed matter to the physics of black holes.
Finding out a viable measurement for the quantum entanglement \cite{qE01} is important, so we consider the mixed bipartite state of the $GCFT_2$ at the boundary. The entanglement entropy is not a proper measure about the entanglement in the mixed states \cite{repl1}, but the entanglement negativity can be used as a measure of the entanglement in the mixed bipartite state configuration of the $GCFT_2$s \cite{repl1}. The approach we have used to calculate entanglement negativity and the entanglement entropy is the replica technique, which has two parts, one is to consider the system manifold in a specific way, and the other is to determine the entanglement based on the reduced density matrix.
To this end the manifold can be considered as an $n$-sheeted surface which involves
$n$ copies of the $GCFT_2$ plane glued together.
After construction of the reduced density matrix for the entangling interval $A$, the replica limit (\ref{EE01}) is applied to find the entanglement entropy of the interval.
By defining the partial transpose of the reduced density matrix in (\ref{red.M02}), another version of the replica limit can be found in (\ref{EN02}) that can be used to compute the entanglement negativity.

The two-point twist correlator (\ref{Q04}) has an important role in computation of the entanglement entropy and the entanglement negativity and is the cornerstone of the correlation functions in other configurations in (\ref{EQ03}), (\ref{4p-temp03}), (\ref{Q-ad02}) and (\ref{4p-2d01}). In other words, the importance of the two-point twist correlator (\ref{Q04}) is not only in the holographic entanglement entropy, but also plays an important role in calculation of the entanglement negativity. We find the appropriate form of the correlator (\ref{Q04}) in the GMMG model. The argument works based on considering the action of the GMMG with a conical singularity. In this approach, we find the entanglement entropy in (\ref{SC01}) relation which is an equation between the conical action and the entanglement entropy. The entanglment entropy is obtained in this line can be found in (\ref{EE-G01}) that is equal to the result of the holographic entanglement entropy that we have found in \cite{Setare3}. This equality can be as an evidence for the validity of the argument. It should be noted the correlator (\ref{Q04}) has two parts that the first one is an $S_{EH}$ contribution which originates from the pure Einstein gravity and its extra terms added in the GMMG. The second contribution in (\ref{Q04}) which is a spinning one originates from the CS-term in the GMMG.

We considered the mixed bipartite state configurations in the cases a single interval, two-adjacent interval and two-disjoint interval. Corresponding to these cases, we considered the field theories that are located at the boundary in three cases, which are in ground state, finite-sized system and finite-temperature state, respectively. One can find the entanglement negativity results of the single interval in three cases in (\ref{EN2-zero}), (\ref{EN2.fin}) and (\ref{EN3-temp}). The differences between these results can be summarized in two parts, which are the different extremal bulk curves and the different extremal boosts. The result of the entanglement negativity related to two-adjacent intervals can be found in (\ref{EN01-adj}), (\ref{EN01-adjFS}) and (\ref{EN01-adjTEMP}).
The result related to the different cases of the two-disjoint intervals can be found in (\ref{EN01-2i0}), (\ref{EN03-2i0}) and (\ref{EN04-2i0}). The coefficients of the extremal bulk curves and the extremal boosts in different cases by two-disjoint and two-adjacent intervals are one-eighth of the central charges. The coefficients in the single interval cases are one-fourth of the central charges.
Such a result in the form of the entanglement negativity is common between TMG and GMMG, but
as we expected, the only difference between the GMMG results and the TMG results is in their central charges that these charges are the coefficients of the Einstein-Hilbert contribution and the spinning contribution in the entanglement negativity.

Another important conclusion that we have obtained is the relation between the holographic entanglement negativity and the entanglement entropy in each cases of the bipartite states. The general form of these relations in the $ GCFT_2$ have been introduced in \cite{Malv}. We extend the results to the GMMG model.
In the single interval cases, the entanglement negativity is three halves of the entanglement entropy in zero-temperature and in the finite-sized state. The only difference is in the finite temperature state that the contribution of the thermal entropy is also included that can be seen in (\ref{3S}). The relation between two types of the entanglement measures in the two-adjacent cases has a same form in each cases of the $GCFT_2$ situations that can be bound in (\ref{EE-EN3}). In these cases, the entanglement negativity is three quarters of the expression in terms of the entanglement entropies of the intervals.

By observing the relation between the entanglement negativity and the entanglement entropy in \cite{repl2} and \cite{calcf}, we arrive at a similar relation between the universal parts of these two types of entanglement measures, as we obtained in (\ref{E&N1}), (\ref{E&N2}) and (\ref{3S}). Of course, it should be noted that the authors of these papers have obtained the entanglement entropy and the entanglement negativity in a single interval state in a relativistic $CFT_2$. The results we have obtained are important because we have been able to produce the universal part of the entanglement measures in relativistic $CFT_2$. This result shows that this part of our results in $GCFT_2$ is produced only through the conformal symmetry. Such a relation between the $CFT$ and the $GCFT$ results can be seen in another work done in \cite{Malv}.

According to the results of the present paper we can say with more sure that
the dual theory of the asymptotically flat spacetime solutions of GMMG is a
$GCFT_2$. The procedure we followed in this paper to find out the entanglement negativity for the mixed states can be used in calculation of the entanglement in two-dimensional condensed matter systems with Galilean conformal symmetry. This can be considered as an open issue that we hope to return to it in the near future.

\section{Acknowledgement}
We thank the anonymous referee for important comments.

\end{document}